\documentclass
[showpacs,floats,amssymb,amsart,amsmath,a4paper,12pt,tightenlines]{iopart}%
\expandafter\let\csname equation*\endcsname\relax
\expandafter\let\csname endequation*\endcsname\relax
\usepackage{amsmath}
\usepackage{epsf}
\usepackage{graphicx}
\usepackage{amsfonts}
\usepackage{amssymb}
\usepackage{dcolumn}%
\providecommand{\U}[1]{\protect\rule{.1in}{.1in}}
\expandafter\let\csname equation*\endcsname\relax
\expandafter\let\csname endequation*\endcsname\relax
\expandafter\let\csname equation*\endcsname\relax
\expandafter\let\csname endequation*\endcsname\relax

\begin{document}

\title{Theory of spin Hall magnetoresistance (SMR) and related phenomena}
\author{Yan-Ting Chen$^{1,2}$, Saburo Takahashi$^{3}$, Hiroyasu Nakayama$^{4}$,
Matthias Althammer$^{5}$, Sebastian T. B. Goennenwein$^{5,6,7}$, Eiji
Saitoh$^{3,8,9,10}$ and Gerrit E. W. Bauer$^{3,8,2}$}

\address{$^{1}$ RIKEN Center for Emergent Matter Science (CEMS), 2-1 Hirosawa, Wako,
Saitama 351-0198, Japan}
\address{$^{2}$ Kavli Institute of NanoScience, Delft University of Technology,
Lorentzweg 1, 2628 CJ Delft, The Netherlands }
\address{$^{3}$ Institute for Materials Research, Tohoku University, Sendai, Miyagi
980-8577, Japan}
\address{$^{4}$ Department of Applied Physics and Physico-Informatics, Keio University, Yokohama 223-8522, Japan}
\address{$^{5}$ Walther-Mei{\ss }ner-Institut, Bayerische Akademie der Wissenschaften,
85748 Garching, Germany}
\address{$^{6}$ Physik-Department, Technische Universit\"{a}t M\"{u}nchen, 85748 Garching, Germany
}
\address{$^{7}$  Nanosystems Initiative Munich (NIM), Schellingstra{\ss}e 4, 80799  M\"{u}nchen, Germany}
\address{$^{8}$ WPI Advanced Institute for Materials Research, Tohoku University,
Sendai 980-8577, Japan}
\address{$^{9}$ CREST, Japan Science and Technology Agency, Tokyo 102-0076, Japan}
\address{$^{10}$ The Advanced Science Research Center, Japan Atomic Energy Agency, Tokai
319-1195, Japan}

\begin{abstract}
We review the so-called spin Hall magnetoresistance (SMR) in bilayers of a
magnetic insulator and a metal, in which spin currents are generated in the
normal metal by the spin Hall effect. The associated angular momentum transfer
to the ferromagnetic layer and thereby the electrical resistance is modulated
by the angle between the applied current and the magnetization direction. The
SMR provides a convenient tool to non-invasively measure the magnetization
direction and spin-transfer torque to an insulator. We introduce the minimal
theoretical instruments to calculate the SMR, i.e. spin diffusion theory and
quantum mechanical boundary conditions. This leads to a small set of
parameters that can be fitted to experiments. We discuss the limitations of
the theory as well as alternative mechanisms such as the ferromagnetic
proximity effect and Rashba spin-orbit torques, and point out new developments.

\end{abstract}
\maketitle




\section{Introduction and motivation}

Spintronics invokes the spin degree of freedom of the electron to improve the
efficiency of and add functionalities to electronic devices
\cite{Bader10,Sinova12}. Here the generation, propagation, processing, and
detection of spin currents form central themes \cite{Maekawa12}. The
spin-orbit coupling (SOC) provides a mechanism to couple the charge and spin
of electrons. A prominent example is the spin Hall effect (SHE), i.e. the
SOC-induced pure spin current transverse to an applied charge current
\cite{Jungwirth12,Hoffmann13,Sinova15}. In bilayers of a normal metal (N: a
non-magnetically ordered metal) and a ferromagnet 
(F, which includes \textquotedblleft{ferrimagnet}\textquotedblright\  
in the present context),
the magnetization dynamics is affected by the SHE spin current via the
spin-transfer torque \cite{Ando08} that can be strong enough to switch the
magnetization \cite{Miron11,Liu12}. On the other hand, the inverse of the SHE
(ISHE), i.e., the pure spin current-driven transverse charge current, is a
convenient tool to detect spin currents generated by spin pumping (SP)
\cite{Saitoh06,Mosendz10-1,Mosendz10-2,Czeschka11} and the spin Seebeck effect
(SSE) \cite{Uchida08,Jaworski10,Uchida10,Weiler12,Uchida14}. Quite some
attention has been recently focused on ferromagnetic insulators (FI) such as
yttrium iron garnet (YIG: Y$_{3}$Fe$_{5}$O$_{12}$) that can be activated
electrically and thermally by attached normal metal contacts. YIG's very low
magnetization damping makes it an attractive material for low-power spin-wave
based interconnects and heat-harvesting devices \cite{WuHoffmann}.

The magnetoresistance (MR) is the ratio of the electrical resistance of a
material with and without an applied magnetic field \cite{Ashcroft76}. In
ferromagnetic conductors the MR is often dominated by effects that are not
caused directly by the applied magnetic field, but rather by the magnetization
orientation as controlled by the applied field. The anisotropic
magnetoresistance (AMR), for example, refers to the phenomenon that the
electrical resistance of ferromagnets depends on the angle between the current
and magnetization directions \cite{Thomson57,McGuire75,Thompson75}. For
applications in magnetic field sensors and data storage devices, the AMR has
been superseded by the giant magnetoresistance (GMR) \cite{Fert08} and
tunneling magnetoresistance \cite{Yuasa08,Moodera10}.

This review addresses a new type of MR discovered in a bilayer made of a metal N
with strong SOC (usually platinum) and an FI (usually YIG), where the
electrical resistance was found to depend on the magnetization of the FI
\cite{Weiler12,Huang12,Nakayama13,Hahn13,Vlietstra13,Althammer13}. Since it
can be explained by the simultaneous action of the SHE and ISHE \cite{Chen13},
it has been dubbed spin Hall magnetoresistance (SMR). This mechanism is
similar to that of the predicted magnetoresistance of the two-dimensional
electron gas induced by magnetic field-induced decoherence of an edge spin
accumulation \cite{Dyakonov07} (the related \textquotedblleft Hanle
effect\textquotedblright\ was observed recently in Pt films under large
magnetic fields \cite{Vevez15}). The SMR can be used to electrically measure
the magnetization direction of an insulator \cite{Saitoh} and helps to
non-invasively access fundamental transport parameters such as the spin Hall
angle, spin diffusion length, and spin-transfer torque (spin-mixing
conductance) at an N$|$FI interface \cite{Liu11}.

The SMR originates from the simultaneous action of two effects, viz. the SHE
and ISHE, that both made an impact on modern spintronics. The SHE causes
current-induced spin-transfer torques without a polarizing ferromagnet
(\textquotedblleft spin-orbit torques\textquotedblright) while the ISHE has
become a standard method of detecting spin currents. However, controversial
issues remain, such as the magnetic proximity effect and the Rashba SOC at the
N$|$FI interface, challenging the applications mentioned above, but also our
understanding of phenomena such as the spin Seebeck effect. The SMR in N$|$FI
bilayers is arguably the most simple observable to detect SOCs: It exists in
bilayers of well-studied and characterized materials, and is a strictly linear
response effect. Hence, it is essential to understand the SMR before more
complicated ones that involve complex devices, difficult materials,
magnetization dynamics, noise or non-linearitis. Many groups have
theoretically and experimentally studied the SMR since its discovery. Hence we
believe that the time has come to consolidate the existing knowledge.

This review is organized as follows: In section~\ref{materials}, we introduce
the basic concepts and prevalent materials necessary to understand and observe
the SMR. In section~\ref{theory}, we describe the diffusion theory for the
(I)SHE with quantum mechanical boundary conditions at interfaces. Experiments
by various groups are discussed in section~\ref{comparison} as well as the
parameters obtained by fitting the theory. In section~\ref{issues}, we address
controversial issues such as magnetic proximity and Rashba SOC at interfaces.
SMR-related new developments are reviewed in section~\ref{developments}. We
present conclusions and an outlook in section~\ref{conclusions}.

\section{Materials and devices}

\label{materials}

In this chapter we review briefly pertinent basic notions in spintronics and
magnetism that are used in the subsequent discussions. It starts with metallic
ferromagnets that have been subject of transport studies for almost two
centuries. Phenomena driven by the SOC in metallic ferromagnets such as the
anisotropic magnetoresistance (AMR) and the anomalous Hall effect (AHE) are
still not fully understood. The spin Hall effect, i.e. the analogue of the AHE
in nonmagnetic metals, has a much shorter history. Its potential to convert
the flows of charge and spin has inspired the whole spintronics community.
Multilayers made of metallic ferromagnets and nonmagnetic metals are known to
display the GMR and TMR as well as spin-transfer torques mediated by the
exchange interaction. More recently, SOC induced spin-transfer torques have
been the center of interest. Here we introduce a few of these notions as
background to the system and phenomenon of interest, i.e. the SMR in N$|$FI bilayers.

\subsection{Metallic ferromagnets, AMR and AHE}

\label{fm_amr_ahe}

Metallic ferromagnets, especially the $3d$ transitions metals and their
alloys, are important in magnetoelectronics due to their high conductivities
and Curie temperatures. Their magnetization is therefore ideal for
non-volatile information storage and electrical readout. The band structure in
a ferromagnet is spin-dependent, as are its transport properties that are well
described by a two parallel current model for majority and minority spins.
Introducing the spin-dependent conductivity $\sigma_{\varsigma F}$
($\varsigma=\uparrow/\downarrow$), an electric field induces the charge
current
\begin{eqnarray}
\vec{j}_{cF}=\vec{j}_{\uparrow F}+\vec{j}_{\downarrow F}=\left(
\sigma_{\uparrow F}+\sigma_{\downarrow F}\right)  \vec{E}\equiv\sigma_{F}%
\vec{E},
\end{eqnarray}
where $\sigma_{F}=\sigma_{\uparrow F}+\sigma_{\downarrow F}$ is the total
conductivity. An applied charge current is accompanied by a spin current
\cite{Mott36}
\begin{eqnarray}
\vec{j}_{sF}=\vec{j}_{\uparrow F}-\vec{j}_{\downarrow F}=\left(
\sigma_{\uparrow F}-\sigma_{\downarrow F}\right)  \vec{E}\equiv P\sigma
_{F}\vec{E},
\end{eqnarray}
with conductance spin polarization
\begin{eqnarray}
P\equiv\frac{\sigma_{\uparrow F}-\sigma_{\downarrow F}}{\sigma_{\uparrow
F}+\sigma_{\downarrow F}}.
\end{eqnarray}
The spin dependence of transport is not easily observed in bulk ferromagnets
directly. The \textquotedblleft{ordinary}\textquotedblright%
\ magnetoresistance, i.e. the dependence of the electrical resistance on an
applied magnetic field exists even for normal metals without SOC through the
Lorentz force \cite{Ashcroft76}. The AMR, i.e. the phenomenon that the
electric resistance depends on the angle between the electric current and the
magnetization vectors, is observed in metallic ferromagnets
\cite{Thomson57,McGuire75}. When current and magnetization direction are given
by unit vectors $\hat{m}$ and $\hat{\jmath}_{c}$, the electric resistance in
an isotropic (or cubic) material reads
\begin{eqnarray}
\rho_{\mathrm{long}}=\rho_{0}+\Delta\rho_{b}\left(  \hat{m}\cdot\hat{\jmath
}_{c}\right)  ^{2},
\end{eqnarray}
where $\Delta\rho_{b}=\rho_{\parallel}-\rho_{\perp}$ with $\rho_{\parallel}$
($\rho_{\perp}$) the resistivity for magnetization parallel (transverse) to
the applied current, and $\rho_{0}$ is an averaged value over all directions
for which different definitions can be found in the literature, for example,
as $\rho_{0}\equiv(\rho_{\parallel}+2\rho_{\perp})/3$ \cite{McGuire75}. A
corollary of the AMR\ is the transverse resistivity $\rho_{\mathrm{trans}}$
also referred to as planar Hall effect. With current direction along $\hat
{x}$
\begin{eqnarray}
\rho_{\mathrm{long}}  &  =\rho_{0}+\Delta\rho_{b}m_{x}^{2}%
,\label{rho_long_amr}\\
\rho_{\mathrm{trans}}  &  =\Delta\rho_{b}m_{x}m_{y}, \label{rho_trans_amr}%
\end{eqnarray}
where $m_{i}$ is the Cartesian $\hat{\imath}$-component of the magnetization
direction unit vector. Higher order contributions to the AMR are found in
single-crystalline thin films \cite{Limmer06}. From the theoretical point of
view, $\Delta\rho_{b}$ can be derived microscopically from the $s$-$d$ model
with a free $s$-electron conduction band and localized $d$-electrons with a
strong exchange interaction and weak SOC. In this model, transport is carried
by the conduction electrons with a contribution to the resistivity from
scattering into localized $d$-states by impurities that depends on the
magnetization direction owing to the SOC. The AMR has been of considerable
interest as a tool to measure the magnetization direction electrically,
thereby serving as a magnetic field sensor \cite{McGuire75}. It still attracts
scientific attention nowadays \cite{Kokado12}, and new regimes are being
opened by studies in ultrathin films \cite{Rijks95,Rijks97,Rowan-Robinson14}.

The \textquotedblleft{ordinary}\textquotedblright\ Hall effect refers to the
transverse charge current normal to both an applied current and external
magnetic field and is named after its discoverer \cite{Hall79}. The AHE, found
by Hall two years later \cite{Hall81}, depends neither on the external
magnetic field nor on the internal dipolar field, but is caused by the SOC or magnetic
orientational disorder (which is generated by SOC) \cite{Nagaosa10}.
Phenomenologically,
\begin{eqnarray}
\vec{j}_{c}^{\mathrm{AHE}}=\theta_{\mathrm{AH}}\hat{m}\times\vec{j}_{c},
\end{eqnarray}
where $\vec{j}_{c}$ is the applied charge current density and $\theta
_{\mathrm{AH}}$ is the anomalous Hall angle, i.e. the ratio between the
transverse and longitudinal conductivities or the slope in plots of the
anomalous Hall conductivity as a function of the longitudinal one.
Controversies about the microscopic mechanism of the AHE, e.g. whether it is
intrinsic or extrinsic, linger, but it appears to be generally accepted that
the extrinsic skew scattering process dominates in the clean regime, while
intrinsic band structure effects and/or extrinsic side jump scattering explain
results for moderately good metals \cite{Nagaosa10,Onoda08,Tian09}. The
anomalous Hall effect can be interpreted in terms of the spin Hall effect
discussed hereafter: In ferromagnets a spin polarized current is
converted into a transverse charge current or voltage by the spin-dependent conductances
as parameterized by the polarization $P$.

\subsection{Spin Hall effect and its inverse in normal metals}

In normal metals without magnetic order, electrons with spin up and spin down
are degenerate and contribute to transport as two channels in parallel.
Nevertheless, significant SOC causes an analogue of the AHE known as the spin
Hall effect (SHE), by which a charge current generates a pure transverse spin
current, i.e., a current of spin angular momentum \cite{Jungwirth12,Sinova15}.
While a spin accumulation can be detected by optical methods at lease in some
systems \cite{Kato04}, we are not aware of experiments that can measure a spin
current directly. However, Onsager reciprocity demands that an inverse SHE
(ISHE) exists, implying that a spin current drives an easily detectable
transverse charge current or voltage \cite{Valenzuela06,Kimura07}.

The SHE and ISHE can be described by
\begin{eqnarray}
\vec{j}_{s\imath}^{\mathrm{SHE}}  &  =\theta_{\mathrm{SH}}\hat{\imath}%
\times\vec{j}_{c},\label{j-she}\\
\vec{j}_{c}^{\mathrm{ISHE}}  &  =\theta_{\mathrm{SH}}\hat{\imath}\times\vec
{j}_{s\imath}. \label{j-ishe}%
\end{eqnarray}
Here $\vec{j}_{s\imath}^{\mathrm{SHE}}/\left\vert \vec{j}_{s\imath
}^{\mathrm{SHE}}\right\vert $ is the direction vector of an SHE spin current
density polarized along $\hat{\imath}$ with modulus $\left\vert \vec
{j}_{s\imath}^{\mathrm{SHE}}\right\vert $. It is driven by the applied charge
current density $\vec{j}_{c}$ and proportional to the spin Hall angle
$\theta_{\mathrm{SH}}$. $\vec{j}_{c}^{\mathrm{ISHE}}$ is the charge current
driven by an $\hat{\imath}$-polarized spin current in $\vec{j}_{s\imath
}/\left\vert \vec{j}_{s\imath}\right\vert $ direction. Note that we define
spin currents in units of Ampere; they can be converted to angular momentum
currents by the factor $\hbar/(2e)$.

\subsection{Metallic multilayers and spin-transfer torques}

Modern crystal growth techniques allow controlled fabrication of multilayers
from various materials with individual film thicknesses of only a few
monolayers. This led to the discovery of the giant magnetoresistance (GMR) in
the current-in-plane (CIP) configuration, i.e., the difference of electric
conductivity between parallel and anti-parallel (magnetic) configurations in
the layered structures \cite{Baibich88,Binash89}. This effect has been
explained by the spin-dependent scattering of electrons at N$|$F interfaces:
in the anti-parallel configuration both spin species are scattered strongly at
opposite interfaces. When an applied magnetic field forces a parallel
configuration one spin channel is short-circuited, leading to a reduced
resistance \cite{Camley89}.

The GMR in the current perpendicular to plane (CPP) configuration is often
larger \cite{Pratt91,Gijs93,Gijs97} and easy to model by one-dimensional spin
diffusion theory \cite{Valet93}. The essential quantity is here the
distributed difference between the effective chemical potentials of electrons
with opposite spins or \textquotedblleft spin accumulation\textquotedblright.
The transfer of spin angular momentum between magnetic layers by an applied
current, i.e., the spin-transfer torque, was predicted
\cite{Slonczewski96,Berger96} and observed
\cite{Tsoi98,Sun99,Myers99,Katine00} in CPP spin valve structures.
Magnetoelectronic circuit theory for magnetic heterostructures with
non-collinear magnetizations provides a theoretical basis to understand the
material dependence of these effects \cite{Brataas06}. By scattering theory,
the spin current $\vec{j}_{s}^{\left(  \mathrm{N}|\mathrm{F}\right)  }$
through an N$|$F interface (on the N side, flowing into F) can be expressed in
terms of the F magnetization $\hat{m}$ and the (vector) spin accumulation
$\vec{\mu}_{sN}$ in N:
\begin{eqnarray}
e\vec{j}_{s}^{\left(  \mathrm{N}|\mathrm{F}\right)  }\left(  \hat{m}\right)
=e\left(  j_{\uparrow}-j_{\downarrow}\right) \hat{m} -G_{r}\hat{m}%
\times\left(  \hat{m}\times\vec{\mu}_{sN}\right)  -G_{i}\left(  \hat{m}%
\times\vec{\mu}_{sN}\right)  , \label{js_interface-0}%
\end{eqnarray}
where $e=-|e|$ is the charge of an electron, and
\begin{eqnarray}
& e j_{\uparrow} =G_{\uparrow}\left[ \left( \mu_{cN}-\mu_{cF}\right)  +\left(
\hat{m} \cdot\vec{\mu}_{sN}-\mu_{sF}\right) /2 \right] ,\\
& e j_{\downarrow} =G_{\downarrow}\left[ \left( \mu_{cN}-\mu_{cF}\right)
-\left( \hat{m} \cdot\vec{\mu}_{sN}-\mu_{sF}\right) /2 \right] .
\end{eqnarray}
are the flows of electrons with spin-up and down electrons along $\hat{m}$
driven by the difference between effective charge chemical potentials in N and
F ($\mu_{cN}-\mu_{cF}$) and the difference between spin accumulations at both
sides of the interface ($\hat{m} \cdot\vec{\mu}_{sN}-\mu_{sF}$). The charge
and spin chemical potentials are related to each other as discussed in
section~\ref{diffusion_theory}, while the spin-dependent conductances at the interface
read
\begin{eqnarray}
& \frac{G_{\uparrow}}{G_{0}} =\sum_{nm}\left[ \delta_{nm}-\left|
r_{nm}^{\uparrow}\right| ^{2}\right] ,\\
& \frac{G_{\downarrow}}{G_{0}} =\sum_{nm}\left[ \delta_{nm}-\left|
r_{nm}^{\downarrow}\right| ^{2}\right] ,
\end{eqnarray}
where $r_{nm}^{\uparrow(\downarrow)}$ is the spin up (down) reflection
coefficient of an electron at the N$|$F interface from transport channel $n$
to channel $m$ in N. Here $G_{0}=e^{2}/h$ is the (single-spin) conductance quantum.

\begin{figure}[ptb]
\includegraphics[width=0.5\textwidth,angle=0]{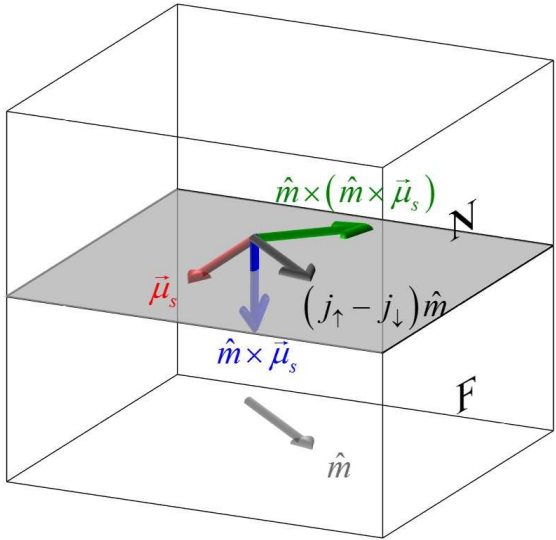}\caption{The spin
current at the N side of an N$|$F interface in (\ref{js_interface-0}) is
the sum of the contributions from the longitudinal conduction electron current
$\left( j_{\uparrow}-j_{\downarrow}\right) \hat{m}$ (that vanishes when F is
an insulator) and the tranverse spin current absorbed at the interface as a
spin-transfer torque. The latter is parametrized by the real and imaginary
parts of the spin-mixing conductance (\ref{def-G}). The vectors indicate
spin current polarizations, while the flow direction is normal to the
interface (from N to F).}%
\label{js-interface-figure}%
\end{figure}

The last two terms in (\ref{js_interface-0}) are spin currents transverse
to the magnetization and parameterized by the spin-mixing conductance at the
interface defined by the elements of the spin-dependent scattering matrix
\begin{eqnarray}
\frac{G_{\uparrow\downarrow}}{G_{0}}=\frac{G_{r}+iG_{i}}{G_{0}}=\sum
_{nm}\left[  \delta_{nm}-r_{nm}^{\uparrow}\left(  r_{nm}^{\downarrow}\right)
^{\ast}\right].  
\label{def-G}
\end{eqnarray}
Figure~\ref{js-interface-figure} illustrates the spin current polarizations at
the interface expressed in (\ref{js_interface-0}). The part of the spin
current with spin polarization along the magnetization orientation in F
component [$\left( j_{\uparrow}-j_{\downarrow}\right) \hat{m}$] can flow in a
metallic F, while the transverse components [$G_{r}\hat{m}\times\left(
\hat{m}\times\vec{\mu}_{sN}\right)  /e$ and $G_{i}\left(  \hat{m}\times
\vec{\mu}_{sN}\right)  /e$] are absorbed at the interface on an (for strong
ferromagnets) atomic length scale and therefore acts as a torque on the
magnetization. The spin-transfer torque at the interface is obtained from
$\vec{j}_{s}^{\left(  \mathrm{N}|\mathrm{F}\right)  }$ by projection
\begin{eqnarray}
\vec{\tau}_{\mathrm{STT}}=\frac{\hbar}{2e}\hat{m}\times\left(  \hat{m}%
\times\vec{j}_{s}^{\left(  \mathrm{N}|\mathrm{F}\right)  }\right)  .
\label{she-stt}%
\end{eqnarray}
(\ref{js_interface-0}) can be used stand-alone for tunnel junctions
or point contacts, or serve as a boundary condition between bulk materials
described by diffusion theory.

\subsection{Spin-orbit torques}

Electron and spin transport in N$|$F bilayers in the current-in-plane (CIP)
configuration attracted attention recently because of the observed large
current-induced spin-transfer torques generated by the SOC
\cite{Gambardella11}. Those spin-orbit torques can result from the spin
currents generated by the SHE in the N layer
\cite{Jungwirth12,Hoffmann13,Sinova15}, and converted to a magnetization
torque by the conventional exchange interaction at the interface. This
contributes a so-called \textquotedblleft{damping-like}\textquotedblright
torque proportional to $G_{r}$ with symmetry identical to the
exchange-mediated term. The magnetic (Gilbert) damping in N$|$F bilayers
subject to an in-plane electric current is modulated by the spin-transfer
torque in (\ref{she-stt}) generated by the direct SHE \cite{Ando08}.
While the SHE generates spin currents, the ISHE in a normal metal is an
efficient detector of spin currents generated, e.g., by spin pumping
\cite{Saitoh06,Mosendz10-1,Mosendz10-2,Czeschka11} or the spin Seebeck effect
\cite{Uchida10}. On the other hand, a Rashba-Edelstein effect at the interface
may generate a spin accumulation that acts directly on the magnetization to
exert a \textquotedblleft{field-like}\textquotedblright\ torque, corresponding
to a finite $G_{i}$ \cite{Edelstein90,Manchon08,Manchon09,Gambardella11}. The
torque on the magnetization induced by the SOC can be strong enough to switch
the magnetization \cite{Miron11,Liu12}, with potential applications in
magnetic storage technology. (\ref{js_interface-0}) may still be used for
the boundary condition of SOC-generated spin currents at N$|$F(I) interface
\cite{Takahashi06}, but the parameters might differ from their exchange-only values.

\subsection{Magnetic insulators}

\label{insulators}

In conventional spintronic devices such as spin valves, metallic ferromagnets
serve as spin emitter (detector) to inject (absorb) spin current into (from)
an adjacent normal metal. The spin transport in this CPP configuration
requires charge flow through the ferromagnet that therefore has to be a metal.
Recently, ferromagnetic insulators such as yttrium iron garnet (YIG) attracted
attention of spintronics researchers \cite{WuHoffmann}. Insulators can
obviously not be used for voltage induced CPP charge flow, but importantly
simplify interpretation in the CIP configuration because parallel charge
transport channels through the magnet do not exist. Ferromagnetic insulators
can have a much better magnetic quality than metals. YIG in particular
features ultralow magnetization damping, which makes it attractive for various
applications \cite{WuHoffmann}.

N$|$F bilayers with an insulating ferromagnet F can be activated electrically
by means of the SHE \cite{Kajiwara10,Collet15}, while the ISHE can be used to
detect magnetization dynamics \cite{Saitoh06,Kajiwara10,Sandweg11}. While
electrical charge transport is suppressed in the CPP configuration, the
thermal activation by a temperature gradient normal to the interface is
possible. A temperature difference at the N$|$F interface generates a spin
current that can be detected by the ISHE in N without complications of
parallel transport in the ferromagnet driven by e.g. the anomalous Nernst
effect. This phenomenon is called (longitudinal) spin Seebeck effect
\cite{Uchida10} and its reciprocal is the spin Peltier effect \cite{Flipse14}.
The theory of the spin Seebeck effect is based on the concepts of thermal spin
pumping and spin-transfer torques in which the spin-mixing conductance and
spin Hall angle are essential parameters \cite{Xiao10,Adachi13}.

\section{Theory}

\label{theory}

In the following we derive in some detail the minimum model for the SMR,
closely following Chen \textit{et al}. \cite{Chen13}.

\subsection{Diffusion theory with spin-flip relaxation but without (I)SHE}

\label{diffusion_theory}

Our theory of the SMR is based on a spin diffusion theory in the limit of weak
SOC. We address the diffusion theories for both ferromagnetic and normal
metals, in which the charge and spin currents are expressed in terms of
gradients of charge and spin accumulations (or spin-dependent electrochemical
potentials and densities). The charge current density is the expectation value
of the current operator $\vec{j}=e\left(  n\vec{v}+\vec{v}n\right)  /2$ in
terms of electron charge $e=-|e|$ , the electrons density $n$, and the
velocity operator $\vec{v}$. For a normal metal with constant density $n_{N}$
and drift velocity $\vec{v}_{N}$,
\begin{eqnarray}
\vec{j}_{cN}=en_{N}\vec{v}_{N}.
\end{eqnarray}
The spin current in the non-relativistic limit
\begin{eqnarray}
\overleftrightarrow{\mathbf{j}_{sN}}=e\left\langle \vec{j}\otimes
\boldsymbol{\vec{\sigma}}+\boldsymbol{\vec{\sigma}}\otimes\vec{j}\right\rangle
/2=\left(  \vec{j}_{sx},\vec{j}_{sy},\vec{j}_{sz}\right)  ^{T}=\left(  \vec
{j}_{s}^{x},\vec{j}_{s}^{y},\vec{j}_{s}^{z}\right)  , \label{sc}%
\end{eqnarray}
is a second-order tensor, where $\boldsymbol{\vec{\sigma}}$ is the vector of
Pauli spin matrices, and $\left\langle \cdots\right\rangle $ denotes an
expectation value. The row vectors $\vec{j}_{si}=en\left\langle \vec
{v}\boldsymbol{\sigma}_{i}+\boldsymbol{\sigma}_{i}\vec{v}\right\rangle /2$ in
(\ref{sc}) are the spin current densities polarized in the
$\hat{\imath}$-direction, while the column vectors $\vec{j}_{s}^{j}%
=en\left\langle v_{j}\boldsymbol{\vec{\sigma}}+\boldsymbol{\vec{\sigma}}%
v_{j}\right\rangle /2$ denote the spin current densities with polarization
$\boldsymbol{\vec{\sigma}}$ flowing in the $\hat{\jmath}$-direction. In
metallic ferromagnets with homogenous texture, the average spin current is
projected along the unit vector of the magnetization direction $\hat{m}$, so
the charge current and spin current tensor read
\begin{eqnarray}
\vec{j}_{cF}  &  =e\left(  n_{\uparrow F}\vec{v}_{\uparrow F}+n_{\downarrow
F}\vec{v}_{\downarrow F}\right)  ,\\
\overleftrightarrow{\mathbf{j}_{sF}}  &  =\vec{j}_{sF}\otimes\hat{m}=\left(
\vec{j}_{\uparrow F}-\vec{j}_{\downarrow F}\right)  \otimes\hat{m}=e\left(
n_{\uparrow F}\vec{v}_{\uparrow F}-n_{\downarrow F}\vec{v}_{\downarrow
F}\right)  \otimes\hat{m},
\end{eqnarray}
where $\vec{j}_{sF}$ is the spin current density direction vector, 
\textquotedblleft$\otimes$\textquotedblright\ denotes that the polarization is locked along $\hat{m}$, 
and $n_{\uparrow/\downarrow}$ and $\vec{v}_{\uparrow/\downarrow}$ are the
spin-dependent electron density and drift velocity of majority/minority
spins in the simple Stoner model. In contrast to the charge current, the
(particle) spin current is not conserved in the presence of SOC and
non-collinear magnetizations, leading to spin-transfer to the lattice or
magnetization, respectively.

In heterostructures, it is useful to express currents in terms of (gradients
of) local quasi-equilibrium thermodynamic parameters such as the
(electro)chemical potential \cite{vanSon87,Valet93,Hershfield97}. At
temperatures well below the magnetic phase transition, transport in
ferromagnetic metals is well described by the two-current model
\cite{Mott36,Fert71}. The spin-dependent electrochemical potentials are
denoted as $\mu_{\varsigma F}$:
\begin{eqnarray}
\mu_{\varsigma F}=e\phi+\delta\mu_{\varsigma F},
\end{eqnarray}
where $\varsigma=\uparrow(\downarrow)$
represents the spin direction (anti-)parallel to the magnetization in the
ferromagnet, and the gradient $-\vec{\nabla}\phi\equiv\vec{E}$ is the external electric
field. As discussed in section~\ref{fm_amr_ahe}, the conductivity in ferromagnets
is spin-dependent and denoted as $\sigma_{\varsigma F}$. Thus we expect that
close to contacts, $\mu_{\uparrow F}\neq\mu_{\downarrow F}$, leading to
spin-dependent diffusion currents
\begin{eqnarray}
\vec{j}_{\varsigma F}=-\frac{\sigma_{\varsigma F}}{e}\vec{\nabla}%
\mu_{\varsigma F}.
\end{eqnarray}
The charge and spin currents now read
\begin{eqnarray}
\vec{j}_{cF}  &  =\vec{j}_{\uparrow F}+\vec{j}_{\downarrow F},\\
\vec{j}_{sF}  &  =\vec{j}_{\uparrow F}-\vec{j}_{\downarrow F},
\end{eqnarray}
and polarization along $\hat{m}$ is presumed. Correspondingly, the charge and
spin electrochemical potentials are defined
\begin{eqnarray}
\mu_{cF}  &  =\frac{\mu_{\uparrow F}+\mu_{\downarrow F}}{2},\\
\mu_{sF}  &  =\mu_{\uparrow F}-\mu_{\downarrow F}.
\end{eqnarray}
With these conventions, we may write Ohm's law in the ferromagnetic metal%
\begin{eqnarray}
\left(
\begin{array}
[c]{c}%
\vec{j}_{cF}\\
\vec{j}_{sF}%
\end{array}
\right)  =\sigma_{F}\left(
\begin{array}
[c]{cc}%
1 & P\\
P & 1
\end{array}
\right)  \left(
\begin{array}
[c]{c}%
-\vec{\nabla}\mu_{cF}/e\\
-\vec{\nabla}\mu_{sF}/(2e)
\end{array}
\right)  . \label{response-0}%
\end{eqnarray}
In a ferromagnet an applied electric field and/or a spin accumulation gradient
generate a charge current as well as a spin current. The charge and spin
electrochemical potentials can be obtained by solving the diffusion equations
\cite{Valet93}.
\begin{eqnarray}
&  \nabla^{2}\mu_{sF} =\frac{\mu_{sF}}{\lambda_{F}^{2}},\label{sdeq-f}\\
&  \nabla^{2}\left(  \mu_{cF}+P\mu_{sF}/2\right)  =0, \label{cdeq-f}%
\end{eqnarray}
where the spin-flip diffusion length $\lambda_{F}=1/\sqrt{\lambda_{\uparrow
F}^{-2}+\lambda_{\downarrow F}^{-2}}$ is expressed in terms of the
spin-diffusion length for each spin $\lambda_{\varsigma F}=\sqrt{D_{\varsigma
F}\tau_{\mathrm{sf},\varsigma F}}$. The spin-dependent charge diffusion
constant $D_{\varsigma F}=\tau_{\varsigma F}v_{\varsigma F}^{2}/3$ depends on
the spin-dependent relaxation time and Fermi velocity, and $\tau
_{\mathrm{sf},\varsigma F}$ is the spin-dependent spin-flip time. With
boundary conditions at contacts, interface and/or deep in the bulk materials,
we may compute $\mu_{cF}$ and $\mu_{sF}$, from which charge and spin currents
are known by (\ref{response-0}).

In normal metals, the electronic structure is spin-degenerate, but spin
accumulations and spin currents can be injected by ferromagnetic contacts or
generated via the SHE. The induced spin accumulations is represented by the
(position dependent) vector
\begin{eqnarray}
\vec{\mu}_{sN}=\left(  \mu_{sx},\,\mu_{sy},\,\mu_{sz}\right)  ^{T}
-\mu_{cN}\hat{1},
\end{eqnarray}
where $\mu_{s\imath}$ represents the $\hat{\imath}$-th Cartesian component.
The charge and spin accumulations obey the diffusion equations
\begin{eqnarray}
\nabla^{2}\mu_{s\imath}  &  =\frac{\mu_{s\imath}}{\lambda^{2}}, \label{sdeq-n}%
\\
\nabla^{2}\mu_{cN}  &  =0. \label{cdeq-n}%
\end{eqnarray}
In the absence of the SHE, charge and spin currents are governed by Fick's
laws \cite{Ashcroft76}
\begin{eqnarray}
\vec{j}_{cN}  &  =-\frac{\sigma_{N}}{e}\vec{\nabla}\mu_{cN},\label{jc-diff}\\
\vec{j}_{s\imath}  &  =-\frac{\sigma_{N}}{2e}\vec{\nabla}\mu_{s\imath}.
\label{js-diff}%
\end{eqnarray}
The difference of the diffusion theory for normal metals with that of
ferromagnets is the arbitrary direction of the spin polarization and the
decoupling between spin and charge when $P\rightarrow0.$

\subsection{Interface boundary conditions}

Solution of the diffusion equations requires boundary conditions for the
currents at surfaces and/or interfaces. Here we disregard interfacial SOC and
proximity effects. This approximation cannot be justified a priori, but
first-principles calculations on N$|$Py$|$N sandwiches [Py: Permalloy
(Ni$_{80}$Fe$_{20}$)] \cite{Liu14,Wang15} can be described by introducing effective
spin-mixing conductances modified by the SOC.

In N$|$FI bilayers charge currents flow in the metal layer parallel to the
applied electric field. The spin currents driven by the SHE generate a spin
accumulation at the interface $\vec{\mu}_{s}^{I}$ that in turn induces a spin
current \cite{Brataas06}
\begin{eqnarray}
e\vec{j}_{s}^{\left(  \mathrm{N}|\mathrm{F}\right)  }\left(  \hat{m}\right)
=-G_{r}\hat{m}\times\left(  \hat{m}\times\vec{\mu}_{sN}^{I}\right)
-G_{i}\left(  \hat{m}\times\vec{\mu}_{sN}^{I}\right)  , 
\label{js_interface}%
\end{eqnarray}
where $G_{r}$ ($G_{i}$) is the real (imaginary) part of the spin-mixing conductance
$G_{\uparrow\downarrow}$ defined by (\ref{def-G}). At the N$|$FI
interface, the reflection coefficients $r_{nm}^{\uparrow(\downarrow
)}=e^{i\delta_{nm}^{\uparrow(\downarrow)}}$ with modulus one and phase
$\delta_{nm}^{\uparrow(\downarrow)}$, resulting in a non-zero value of
$G_{\uparrow\downarrow}$:
\begin{eqnarray}
\frac{G_{\uparrow\downarrow}}{G_{0}}=N_{\mathrm{Sh}}-\sum_{n}r_{nm}^{\uparrow
}\left(  r_{nm}^{\downarrow}\right)  ^{\ast}=N_{\mathrm{Sh}}-\sum
_{nm}e^{i\left(  \delta_{nm}^{\downarrow}-\delta_{nm}^{\uparrow}\right)  },
\end{eqnarray}
where $N_{\mathrm{Sh}}$ is the number of transport channels (per unit area) at
the Fermi energy, \textit{i.e.} the Sharvin conductance (for one spin) in N.
Therefore,
\begin{eqnarray}
\frac{G_{r}}{G_{0}}\leq2N_{\mathrm{Sh}};\;\frac{\left\vert G_{i}\right\vert
}{G_{0}}\leq N_{\mathrm{Sh}}.
\end{eqnarray}
The original circuit theory assumes that the nodes are in local equilibrium.
In highly conductive systems, the interface conductances have to be corrected
for the electron drift by subtracting spurious Sharvin conductance (Schep
correction) \cite{Schep97,Bauer03}. The corrected mixing conductance
$\tilde{G}_{r}$ reads
\begin{eqnarray}
\frac{1}{\tilde{G}_{r}/G_{0}}=\frac{1}{G_{r}/G_{0}}-\frac{1}{2N_{\mathrm{Sh}}%
}.
\end{eqnarray}

\subsection{SMR}

\label{theory-smr}\begin{figure}[ptb]
\includegraphics[width=0.5\textwidth,angle=0]{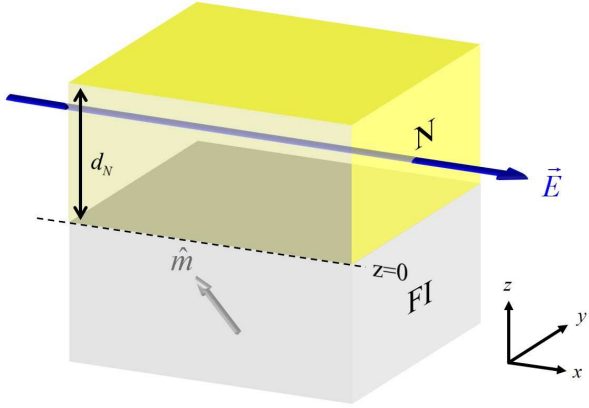} \caption{The
N$|$FI bilayer system with FI a ferromagnetic insulator and N a normal metal.}%
\label{bilayer}%
\end{figure}Here we derive an expression of the SMR for the bilayer in
figure~\ref{bilayer}. We can generalize Ohm's law for metals with a weak SHE,
i.e., a small $\theta_{\mathrm{SH}}$, as a relation between thermodynamic
driving forces and currents that reflects Onsager's reciprocity by the
symmetry of the response matrix~\cite{Takahashi06}:
\begin{eqnarray}
\fl\left(
\begin{array}
[c]{c}%
\vec{j}_{cN}\\
\vec{j}_{sx}\\
\vec{j}_{sy}\\
\vec{j}_{sz}%
\end{array}
\right)  =\sigma_{N}\left(
\begin{array}
[c]{cccc}%
1 & \theta_{\mathrm{SH}}\hat{x}\times & \theta_{\mathrm{SH}}\hat{y}\times &
\theta_{\mathrm{SH}}\hat{z}\times\\
\theta_{\mathrm{SH}}\hat{x}\times & 1 & 0 & 0\\
\theta_{\mathrm{SH}}\hat{y}\times & 0 & 1 & 0\\
\theta_{\mathrm{SH}}\hat{z}\times & 0 & 0 & 1
\end{array}
\right)  \left(
\begin{array}
[c]{c}%
-\vec{\nabla}\mu_{cN}/e\\
-\vec{\nabla}\mu_{sx}/(2e)\\
-\vec{\nabla}\mu_{sy}/(2e)\\
-\vec{\nabla}\mu_{sz}/(2e)
\end{array}
\right)  , \label{response}%
\end{eqnarray}
where $\vec{\mu}_{sN}=\left(  \mu_{sx},\mu_{sy},\mu_{sz}\right)  ^{T}-\mu
_{cN}\hat{1}$ is the spin accumulation, \textit{i.e.} the spin-dependent
chemical potential relative to the charge chemical potential $\mu_{cN}=e\phi$,
$\sigma_{N}$ is the electric conductivity, $\theta_{\mathrm{SH}}$ is the spin
Hall angle, and \textquotedblleft$\times$\textquotedblright\ denotes the
vector cross product operating on the gradients of the spin-dependent chemical
potentials. The SHE is represented by the \textquotedblleft
lower\textquotedblright\ non-diagonal elements that generate the spin currents
in the presence of an applied electric field, in the following chosen to be in
the $\hat{x}$-direction $\vec{E}=E_{x}\hat{x}=-\hat{x}\partial_{x}\mu_{cN}/e$.
The ISHE is governed by elements \textquotedblleft above\textquotedblright%
\ the diagonal that connect the gradients of the spin accumulations to the
charge current density. (\ref{response}) is valid for cubic lattices such
as Pt.

For films with thickness $d_{N}$ in the $\hat{z}$-direction as shown in
figure~\ref{bilayer}, the general solution of the spin diffusion (\ref{sdeq-n}) reads
\begin{eqnarray}
\vec{\mu}_{sN}\left(  z\right)  =\vec{A}e^{-z/\lambda}+\vec{B}e^{z/\lambda},
\label{mus_0}%
\end{eqnarray}
where the constant column vectors $\vec{A}$ and $\vec{B}$ are determined by
the boundary conditions at the interfaces. We do not have to consider the
charge accumulation since YIG is an insulator and the SOC does not generate
any transverse charge current (via the AHE) as long as N remains non-magnetic.

According to (\ref{response}), the spin current in N consists of
conventional diffusion and spin Hall drift contributions. Assuming
translational invariance in the $x$-$y$ plane, we focus on the spin current
density flowing in the $\hat{z}$-direction
\begin{eqnarray}
\vec{j}_{s}^{z}(z)=-\frac{\sigma_{N}}{2e}\partial_{z}\vec{\mu}_{sN}%
-j_{s0}^{\mathrm{SH}}\hat{y}, \label{js-flows-z}%
\end{eqnarray}
where $j_{s0}^{\mathrm{SH}}=\theta_{\mathrm{SH}}\sigma_{N}E_{x}$ is the bare
spin Hall current, \textit{i.e.}, the spin current generated directly by the
SHE. The boundary conditions require that $\vec{j}_{s}^{z}(z)$ is continuous
at the interfaces $z=d_{N}$ and $z=0$. The spin current density at a vacuum
interface ($z=d_{N}$) vanishes, $\vec{j}_{s}^{z}(d_{N})=0$ 
while at the magnetic interface ($z=0$), it is governed by the spin accumulation 
and spin-mixing conductance according to (\ref{js_interface}),
$\vec{j}_{s}^{z}(0)=-\vec{j}_{s}^{\left(  \mathrm{N}|\mathrm{F}\right)}$, where the minus sign is due to that $\vec{j}_{s}^{\left(  \mathrm{N}|\mathrm{F}\right)}$ is flowing from N to F, i.e., in the $-\hat{z}$-direction. 

With these boundary conditions we can determine the coefficients $\vec{A}$ and
$\vec{B}$ for the bilayer, which leads to the spin accumulation
\begin{eqnarray}
\vec{\mu}_{sN}(z)=-\hat{y}\mu_{s}^{0}\frac{\sinh\frac{2z-d_{N}}{2\lambda}%
}{\sinh\frac{d_{N}}{2\lambda}}+\vec{j}_{s}^{\left(  \mathrm{N}|\mathrm{F}%
\right)  }\left(  \hat{m}\right)  \frac{2e\lambda}{\sigma_{N}}\frac{\cosh
\frac{z-d_{N}}{\lambda}}{\sinh\frac{d_{N}}{\lambda}},
\end{eqnarray}
where $\mu_{s}^{0}\equiv\vec{\mu}_{sN}(0)=(2e\lambda/\sigma_{N})j_{s0}%
^{\mathrm{SH}}\tanh\left[  d_{N}/\left(  2\lambda\right)  \right]  $ is the
spin accumulation at the interface in the absence of spin-transfer,
\textit{i.e.}, when $G_{\uparrow\downarrow}=0$.

Using (\ref{js_interface}), we find the spin accumulation
\begin{eqnarray}
\fl\frac{\vec{\mu}_{sN}(z)}{\mu_{s}^{0}}  &  =-\hat{y}\frac{\sinh
\frac{2z-d_{N}}{2\lambda}}{\sinh\frac{d_{N}}{2\lambda}}\nonumber\\
\fl  &  +\left[  \hat{m}\times\left(  \hat{m}\times\hat{y}\right)
\operatorname{Re}+\left(  \hat{m}\times\hat{y}\right)  \operatorname{Im}%
\right]  \frac{2\lambda G_{\uparrow\downarrow}}{\sigma_{N}+2\lambda
G_{\uparrow\downarrow}\coth\frac{d_{N}}{\lambda}}\frac{\cosh\frac{z-d_{N}%
}{\lambda}}{\sinh\frac{d_{N}}{\lambda}}, \label{musgen}%
\end{eqnarray}
that leads to the distributed spin current in N
\begin{eqnarray}
\fl\frac{\vec{j}_{s}^{z}(z)}{j_{s0}^{\mathrm{SH}}}  &  =\hat{y}\frac
{\cosh\frac{2z-d_{N}}{2\lambda}-\cosh\frac{d_{N}}{2\lambda}}{\cosh\frac{d_{N}%
}{2\lambda}}\nonumber\\
\fl  &  -\left[  \hat{m}\times\left(  \hat{m}\times\hat{y}\right)
\operatorname{Re}+\left(  \hat{m}\times\hat{y}\right)  \operatorname{Im}%
\right]  \frac{2\lambda G_{\uparrow\downarrow}\tanh\frac{d_{N}}{2\lambda}%
}{\sigma_{N}+2\lambda G_{\uparrow\downarrow}\coth\frac{d_{N}}{\lambda}}%
\frac{\sinh\frac{z-d_{N}}{\lambda}}{\sinh\frac{d_{N}}{\lambda}}.
\end{eqnarray}

\begin{table}[ptb]
\caption{Normalized $j_{sy}$, $j_{sx}$, $\mu_{sy}$, and $\mu_{sx}$ at the top
($z=d_{N}$) and bottom ($z=0$) of N for magnetizations $\hat{m}=\hat{y}$
(completely reflecting), $\hat{m}=\left(  \hat{x}+\hat{y}\right)  /\sqrt{2}$,
and $\hat{m}=\hat{x}$ (maximally absorbing). At the top (contacted to vacuum),
the values do not depend on the magnetization; while at the bottom (contacted
to FI), the values are strongly affected by the orientation of magnetization.
We adopt the parameters $d_{N}=12$~nm, $\rho=8.6\times10^{-7}%
\operatorname{\Omega}\operatorname{m}$, $\lambda=1.5$~nm, and $G_{r}%
=5\times10^{14}\operatorname{\Omega}^{-1}\operatorname{m}^{-2}$.}%
\label{tb-0}%
\begin{indented}
\item[]\begin{tabular}[c]{ccccc}
\br & $j_{sy}(d_{N})/j_{s0}^{\mathrm{SH}}$ & $j_{sx}(d_{N})/j_{s0}^{\mathrm{SH}}$ &
$\mu_{sy}(d_{N})/\mu_{s}^{0}$ & $\mu_{sx}(d_{N})/\mu_{s}^{0}$ \\
& 0& 0& -1& 0\\
\mr
& $j_{sy}(0)/j_{s0}^{\mathrm{SH}}$ & $j_{sx}(0)/j_{s0}^{\mathrm{SH}}$ &
$\mu_{sy}(0)/\mu_{s}^{0}$ & $\mu_{sx}(0)/\mu_{s}^{0}$ \\
$\hat{m}=\hat{y}$& 0& 0& 1& 0\\
$\hat{m}=\left(\hat{x}+\hat{y}\right)/\sqrt{2}$& -0.28& 0.28& 0.72& 0.28\\
$\hat{m}=\hat{x}$& -0.56& 0& 0.44& 0\\
\br
\end{tabular}
\end{indented}
\end{table}

First we have a look at the spatial dependences of the spin current and spin
accumulation. According to first principles calculations~\cite{Jia11},
$\left\vert G_{i}\right\vert $ is at least one order of magnitude smaller than
$G_{r}$ for YIG, so $G_{i}=0$ appears to be a good first approximation. In
this limit, the normalized components of spin current ($j_{sx}=\vec{j}_{s}%
^{z}\cdot\hat{x}$ and $j_{sy}=\vec{j}_{s}^{z}\cdot\hat{y}$) and spin
accumulation ($\mu_{sx}$ and $\mu_{sy}$) at the top ($z=d_{N}$) and bottom
($z=0$) of N for different magnetizations are shown in table~\ref{tb-0}. When
the magnetization of F is along $\hat{y}$, the spin current at the N$|$F
interface ($z=0$) vanishes just as for the vacuum interface. By rotating the
magnetization from $\hat{y}$ to $\hat{x}$, the spin current at the N$|$F
interface and the torque on the magnetization is activated, while the spin
accumulation is dissipated correspondingly. The $x$-components of both spin
accumulation and spin current vanish when the magnetization is along $\hat{x}$
and $\hat{y}$, and are largest at $\left(  \hat{x}+\hat{y}\right)  /\sqrt{2}$.
This behavior agrees with the relations between spin current and spin-transfer
torque at a N$|$F interface discussed above.

The ISHE drives a charge current in the $x$-$y$ plane by the diffusion spin
current component flowing along the $\hat{z}$-direction. The total
longitudinal (along $\hat{x}$) and transverse or Hall (along $\hat{y}$) charge
currents become
\begin{eqnarray}
\fl \frac{j_{c,\mathrm{long}}(z)}{j_{c}^{0}}  &  =1+\theta_{\mathrm{SH}}%
^{2}\left[  \frac{\cosh\frac{2z-d_{N}}{2\lambda}}{\cosh\frac{d_{N}}{2\lambda}%
}+\left(  1-m_{y}^{2}\right)  \operatorname{Re}\frac{2\lambda G_{\uparrow
\downarrow}\tanh\frac{d_{N}}{2\lambda}}{\sigma_{N}+2\lambda G_{\uparrow
\downarrow}\coth\frac{d_{N}}{\lambda}}\frac{\sinh\frac{z-d_{N}}{\lambda}%
}{\sinh\frac{d_{N}}{\lambda}}\right]  ,\\
\fl \frac{j_{c,\mathrm{trans}}(z)}{j_{c}^{0}}  &  =\theta_{\mathrm{SH}}%
^{2}\left(  m_{x}m_{y}\operatorname{Re}-m_{z}\operatorname{Im}\right)
\frac{2\lambda G_{\uparrow\downarrow}\tanh\frac{d_{N}}{2\lambda}}{\sigma
_{N}+2\lambda G_{\uparrow\downarrow}\coth\frac{d_{N}}{\lambda}}\frac
{\sinh\frac{z-d_{N}}{\lambda}}{\sinh\frac{d_{N}}{\lambda}},
\end{eqnarray}
where $j_{c}^{0}=\sigma_{N}E_{x}$ is the charge current driven by the external
electric field.

The charge current vector is usually expressed in terms of the longitudinal
and transverse (Hall) resistivities. Averaging the electric currents over the
film thickness $z$ and expanding the longitudinal resistivity or current in
the ($x$-)direction of the applied field to leading order in $\theta
_{\mathrm{SH}}^{2}$, we obtain
\begin{eqnarray}
\rho_{\mathrm{long}}  &  =\sigma_{\mathrm{long}}^{-1}=\left(  \frac
{\overline{j_{c,\mathrm{long}}}}{E_{x}}\right)  ^{-1}\approx\rho+\Delta
\rho_{0}+\Delta\rho_{1}\left(  1-m_{y}^{2}\right)  ,\label{rho_long_smr}\\
\rho_{\mathrm{trans}}  &  =-\frac{\sigma_{\mathrm{trans}}}{\sigma
_{\mathrm{long}}^{2}}\approx-\frac{\overline{j_{c,\mathrm{trans}}}/E_{x}%
}{\sigma_{N}^{2}}=\Delta\rho_{1}m_{x}m_{y}+\Delta\rho_{2}m_{z},
\label{rho_trans_smr}%
\end{eqnarray}
where
\begin{eqnarray}
\fl \frac{\Delta\rho_{0}}{\rho}  &  =-\theta_{\mathrm{SH}}^{2}\frac{2\lambda
}{d_{N}}\tanh\frac{d_{N}}{2\lambda},\label{rho-0}\\
\fl \frac{\Delta\rho_{1}}{\rho}  &  =\theta_{\mathrm{SH}}^{2}\frac{\lambda
}{d_{N}}\operatorname{Re}\frac{2\lambda G_{\uparrow\downarrow}\tanh^{2}%
\frac{d_{N}}{2\lambda}}{\sigma_{N}+2\lambda G_{\uparrow\downarrow}\coth
\frac{d_{N}}{\lambda}}\approx\theta_{\mathrm{SH}}^{2}\frac{\lambda}{d_{N}%
}\frac{2\lambda G_{r}\tanh^{2}\frac{d_{N}}{2\lambda}}{\sigma_{N}+2\lambda
G_{r}\coth\frac{d_{N}}{\lambda}},\label{rho-1}\\
\fl \frac{\Delta\rho_{2}}{\rho}  &  =-\theta_{\mathrm{SH}}^{2}\frac{\lambda
}{d_{N}}\operatorname{Im}\frac{2\lambda G_{\uparrow\downarrow}\tanh^{2}%
\frac{d_{N}}{2\lambda}}{\sigma_{N}+2\lambda G_{\uparrow\downarrow}\coth
\frac{d_{N}}{\lambda}}\approx-\theta_{\mathrm{SH}}^{2}\frac{\lambda}{d_{N}%
}\frac{2\lambda\sigma_{N}G_{i}\tanh^{2}\frac{d_{N}}{2\lambda}}{\left(
\sigma_{N}+2\lambda G_{r}\coth\frac{d_{N}}{\lambda}\right)  ^{2}},
\label{rho-2}%
\end{eqnarray}
and $\rho=\sigma_{N}^{-1}$ is the intrinsic electric resistivity of the bulk
normal metal. The last approximations requires $G_{r}\gg G_{i}$. We may not
conclude from $\Delta\rho_{0}<0$ that SOC reduces the global resistivity,
since the bulk $\rho$ itself is increased by spin-flip scattering in the bulk
that is not explicitly treated here. For general magnetization and current
directions (with film in the $x$-$y$ plane) we can summarize the angle
dependence as \cite{Sinova15}%
\begin{eqnarray}
\rho_{\mathrm{long}}  &  =\rho+\Delta\rho_{0}+\Delta\rho_{1}-\Delta\rho
_{1}\left[  \hat{m}\cdot\left(  \hat{\jmath}_{c}\times\hat{z}\right)  \right]
^{2},\\
\rho_{\mathrm{trans}}  &  =-\Delta\rho_{1}\left(  \hat{m}\cdot\hat{\jmath}%
_{c}\right)  \left[  \hat{m}\cdot\left(  \hat{\jmath}_{c}\times\hat{z}\right)
\right]  +\Delta\rho_{2}\left(  \hat{m}\cdot\hat{z}\right)  .
\end{eqnarray}
(\ref{rho_long_smr}-\ref{rho-2}) are the main results of the SMR
model that can be compared with and fitted to experiments. We also note that
the magnetization orientation dependence of the resistivity derived here holds
for an isotropic N. In the presence of crystalline anisotropy, the angle
dependence may involve higher harmonics \cite{Limmer06}.

\section{Comparison with experiments}

\label{comparison}

\begin{table}[ptb]
\caption{Transport parameters $(\theta_{\mathrm{SH}},\,\lambda,\,G_{r})$ in
N$|$FI bilayers obtained from experiments: A$=$SMR, B$=$SSE, C$1=$SP$+$FMR,
and C$2=$SP$+$ISHE. $^{\ast}$ labeled fits employed parameters from other
sources.}%
\label{tb1}%
\begin{indented}
\item[]\begin{tabular}[c]{ccccccc}
\br FI & N & $\theta_{\mathrm{SH}}$\thinspace(\%) & $\lambda$\thinspace(nm) &
$G_{r}/G_{0}$\thinspace($10^{19}$m$^{-2}$) & method & Ref.\\
\mr YIG & Pt & 11 & 1.5 & 1 & A & \cite{Althammer13}\\
&  & 3 & 2.5 & 1 & A & \cite{Nakayama13}\\
&  & 8 & 1.5 & 0.3 & A & \cite{Vlietstra13}\\
&  & 8 & 1.2 & 1.1 & A & \cite{Vlietstra13-2}\\
&  & 11 & 1.5 & 1 & A$+$B$+$C$2$ & \cite{Weiler13}\\
&  & 3 &  & 0.16 & A+C$2^{\ast}$ & \cite{Hahn13}\\
&  & 5 & 2 & 0.26 & C$2^{\ast}$ & \cite{Hahn13-2}\\
&  &  &  & 48 & C$1$ & \cite{Rezende13}\\
&  &  &  & 3.43 & C$1$ & \cite{Jungfleisch13}\\
&  &  &  & 0.13 & C$2^{\ast}$ & \cite{Qiu13}\\
&  &  1.3 & 2.5 & & B$^{\ast}$ &\cite{Qu14}\\
& Ta & -2 & 1.8 & 0.11 & A$+$C$2$ & \cite{Hahn13}\\
& & -0.14 & 1.7 & & B$^{\ast}$ &\cite{Qu14}\\
& Au & 0.16 & 40 &  & B & \cite{Qu13}\\
&  &  0.3 & 9.5 & & B$^{\ast}$ &\cite{Qu14}\\
&  &  &  & 0.12 & C$1$ & \cite{Heinrich11}\\
&  &  &  & 0.5 & C$1$ & \cite{Burrowes12}\\
& W & -0.43 & 1.5 & & B$^{\ast}$ & \cite{Qu14}\\
CFO(001) & Pt &  &  & 0.65 & A$^{\ast}$ & \cite{Isasa14}\\
CFO(111) & Pt &  &  & 0.39 & A$^{\ast}$ & \cite{Isasa14}\\
\mr
$G_{0}=3.87\times10^{-5}\Omega^{-1}$\\
\br
\end{tabular}
\end{indented}
\end{table}

The SMR theory sketched above leads to simple analytic forms
~(\ref{rho_long_smr}-\ref{rho-2}) that predict the magnetization dependence of
the electric resistivity of the bilayers in terms of one sample parameter,
i.e. the thickness $d_{N}$, and four material parameters: the conductivity
$\sigma_{N}$, spin Hall angle $\theta_{\mathrm{SH}}$, spin diffusion length
$\lambda$, and spin-mixing conductance $G_{\uparrow\downarrow}=G_{r}+iG_{i}$,
where $G_{i}$ is believed small and treated \textit{a posteriori} by
perturbation theory. The parameters may be fitted to SMR observations and
other independent experiments such as ferromagnetic resonance (FMR) on the
same bilayers.

While the SMR was discovered and filed as a patent as a remote detector of the
magnetization direction of a ferromagnetic insulator a few years ago
\cite{Saitoh}, the effect was first published in the supplemental material of
\cite{Weiler12}. Initially only the in-plane magnetization dependence was
measured in Pt$|$YIG, which is phenomenologically identical to the
conventional AMR. Weiler \textit{et al}. therefore suspected that a
proximity-induced ferromagnetic Pt layer could play a role \cite{Weiler12},
which implied that the spin Seebeck effect in Pt$|$YIG could be caused by an
anomalous Nernst effect \cite{Huang12}.

Subsequent experimental efforts addressed the dependence of the SMR when the
magnetization is forced out of the interface plane, which revealed that the
SMR is not consistent with a conventional AMR in a polycrystalline bulk
material \cite{Nakayama13,Althammer13}. While the SMR is proportional to
$m_{y}^{2}$ according to (\ref{rho_long_smr}) and
(\ref{rho_trans_smr}), the conventional AMR depends on $m_{x}^{2}$ according
to (\ref{rho_long_amr}) and (\ref{rho_trans_amr}). These conclusions
were confirmed by other experiments
\cite{Vlietstra13,Vlietstra13-2,Hahn13,Lin13,Lin14} also for bilayers with
other metals such as Ta \cite{Hahn13} and Pd \cite{Lin13,Lin14}. The current
level required for SMR experiments can be kept low, but Joule heating
artifacts can be excluded by using lock-in techniques or symmetry arguments
\cite{Vlietstra14,Schreier13}, as well.

Measurements on Pt$|$Cu$|$YIG and Pt$|$Au$|$YIG provide more tests for the
plausibility of a proximity effect. Cu and Au are diamagnetic and not
susceptible to a ferromagnetic proximity effect. The persistence of an MR in
the above trilayers can then be interpreted in terms of the spin Hall effect
in Pt \cite{Nakayama13,Althammer13} when the parallel conductance channels of
the high-mobility Cu and Au spacer layers are taken into account, as shown in
the supplementary material of \cite{Nakayama13}.

The essential material-dependent parameters $(\theta_{\mathrm{SH}}%
,\,\lambda,\,G_{\uparrow\downarrow})$ are assumed to be intensive, i.e. to not
depend on the layer thicknesses, which is a debatable approximation (see
section~\ref{interface-roughness}). They can then be determined by fitting
(\ref{rho-1}) derived above to the experimental SMR for various $d_{N}$.
The spin Hall angle and spin diffusion length both reflect the SOC and might
be correlated with each other \cite{Schreier}, but we treat them here as
independent phenomenological parameters \cite{Chen13}. The extracted
parameters by different groups for the N$\vert$FI
\cite{Huang12,Nakayama13,Althammer13}, and those from earlier experiments e.g. FMR on
metallic bilayers \cite{Liu11} are of the same order of magnitude. The
transport parameters have been extracted from experiments of spin pumping
detected by the ISHE in conjunction with SMR results on bilayers of Pt$|$YIG
and Ta$|$YIG by Hahn \textit{et al}. \cite{Hahn13}: the thickness dependence
of the ISHE voltage was used to fit the spin diffusion length $\lambda$ using
the expressions from \cite{Castel12}. The spin-mixing conductances can then be
accurately fitted to the magnetoresistance because the ratio of the square of
the spin pumping-ISHE voltage and the SMR does not depend on $\theta
_{\mathrm{SH}}$. They confirmed a different sign for Pt ($\theta_{\mathrm{SH}%
}>0$) and Ta ($\theta_{\mathrm{SH}}<0$) \cite{Schreier15}. A systematic study
including SSE, spin pumping (probed by the ISHE and FMR spectral broadening),
and the SMR \cite{Weiler13} led to a single set of parameters for these three
effects, confirming the presumed identical physical origin. The real part of
spin-mixing conductance appears to be in the interval $0.5\times
10^{19}\mathrm{m}^{-2}\leq G_{r}\leq1.5\times10^{19}\mathrm{m}^{-2}$ for
various samples \cite{Weiler13}. The large value of the mixing conductance
with YIG was predicted by first-principle calculations \cite{Jia11} and
confirmed earlier by FMR experiments \cite{Burrowes12}. Parameters fitted from
the SMR are listed in table~\ref{tb1} and compared with those obtained by
other methods for N$|$FI heterostructures.

Besides the planar Hall effect (the transverse corollary of a longitudinal
MR), a Hall voltage that scales linear with the $z$-component of the
magnetization (normal to the film) has been observed, i.e. an anomaloul
Hall-like effect that can be explained by the SMR mechanism when the imaginary
part of the spin-mixing conductance $G_{i}$ in (\ref{rho-2}) is
significant. Combining experiments on the magnetization direction and film
thickness dependences, a value for $G_{i}$ can be extracted
\cite{Althammer13,Vlietstra13-2} that agrees with first principle calculations
\cite{Jia11}. Alternative explanations in terms of a conventional AHE in a
magnetized monolayer of Pt or surface roughness, however, cannot be excluded
as discussed in section~\ref{issues}.

The SMR has been observed in bilayers made from other metals than Pt such as
Ta, Pd, W, Nb, and Rh \cite{Hahn13,Yang14,Lin13,Lin14,Qu14,Han15,Shang15-2},
and other magnetic insulators besides YIG \cite{Althammer13,Isasa14,Han14}. In
Pt$|$CFO (CFO: CoFe$_{2}$O$_{4}$) bilayers, the spin-mixing conductance was
found to strongly depend on the crystallographic interface orientation,
possibly due to different densities of magnetic moments \cite{Isasa14}, which
should be confirmed by spin pumping and SSE experiments and first principle
calculations. The dependence of the SMR on the magnetic ordering at the
interface provides a convenient tool to probe the interface magnetization of
the FI, which can be very different from the bulk magnetization \cite{Isasa15-2}.
The SMR with Pt on the antiferromagnet SrMnO$_{3}$ was found to be finite only
when a net magnetization was induced by an applied magnetic field
\cite{Han14}. Since a current-induced spin-transfer torque has been predicted
for antiferromagnets \cite{Takei14,Cheng14}, we expect an SMR also for
unpolarized, single-domain antiferromagnets.

The SMR has been observed for Pt$|$Fe$_{3}$O$_{4}$ bilayers at low
\cite{Althammer13} and room temperatures \cite{Ding14}. Fe$_{3}$O$_{4}$ is not
an insulator, but a bad metal with resistivity about five orders of magnitude
larger than that of Pt$|$Fe$_{3}$O$_{4}$ and a possible AMR in the ferrite is
likely to be shunted \cite{Ding14}. The SMR is also observed in metallic
layered systems, even though its interpretation is more difficult due to
currents in the ferromagnet and possibly more serious proximity effects. MR
measurements in Pt$|$Co$|$Pt and Pt$|$Py$|$Pt trilayers \cite{Kobs11,Lu13-1}
found a magnetization dependence that is a combination of the AMR and what we
now call SMR phenomenology, revealing that the SMR could be present also in
all-metallic samples. In Ta$|$Co, Pt$|$Co and W$|$Pt, a difference in the
resistance by, respectively, two (Ta$|$Co) and three (Pt$|$Co and W$|$Pt)
orders of magnitude smaller than AMR$+$SMR was found in N$|$FM bilayers for
magnetizations parallel and antiparallel to $\hat{\jmath}_{c}\times\hat{z}$
\cite{Avci15,Avci15-2}. Indeed, in ferromagnetic metals (in contrast to
insulators) the reflection of the spin Hall current at the interface is
incomplete and differs when parallel and antiparallel to the magnetization
\cite{Avci15}. To capture this contribution, one may have to extend the SMR
model by including the diffusive transport in F and fitting the boundary
condition with (\ref{js_interface-0}) rather than with (\ref{js_interface}). 
It was reported that in Ga$_{0.91}$Mn$_{0.09}%
$As$|$Ga$_{0.97}$Mn$_{0.03}$As (FM$|$NM), this difference in MR can be in the
same order of the SMR$+$AMR \cite{Olejnik15}. An SMR in metallic bilayers
W$|$CoFeB has been reported to be an order of magnitude larger than that
in N$|$FI, which was interpreted in terms of a spin Hall angle $\theta
_{\mathrm{SH}}\approx0.2$ for $\beta$-tungsten \cite{Kim15,Cho15}. A complete
model for the spin-orbit torques in bilayers should explain the observed
correlation between the SMR amplitude and current-induced magnetization
switching in these bilayers \cite{Cho15}.

\section{Issues}

\label{issues} Here we address controversial and unsolved issues concerning
the interpretation of experiments in terms of the SMR mechanism.

\subsection{Magnetic proximity effects}

\label{proximity}

As mentioned above, the observation of a magnetoresistance in Pt on a magnetic
insulator could in principle be caused by the equilibrium magnetic proximity
effect (MPE). Pt and Pd have Stoner-enhanced magnetic susceptibilities and are
therefore \textquotedblleft{almost}\textquotedblright\ ferromagnetic
\cite{Buschow03}. For instance, magnetic impurities in these metals have a
strongly enhanced magnetic moment. The MPE is well established in bilayers
made of a normal metal and a ferromagnetic metal. X-ray magnetic circular
dichroism (XMCD) \cite{Wende04} detected a large polarization at the interface
of Pt on a magnetic metal \cite{Ruegg91,Antel99,Wilhelm00,Wilhelm01,Meier11}%
\ as compared to less-susceptible metals. For example, the MPE in Au$|$Co is
about one order of magnitude smaller than that in Pt$|$Co \cite{Wilhelm04}.
XMCD studies on Pt$|$YIG led to conflicting results \cite{Geprags12,Lu13}.
Theoretically, spin-polarized density functional calculations for Pt$|$YIG and
Au$|$YIG found the Pt spin polarization in Pt$|$YIG to be stronger than that
of Au in Au$|$YIG \cite{Qu13}. However, since YIG is a ferrimagnet with a much
smaller saturation magnetization, the MPE in Pt$|$YIG may be expected be
smaller than that for strongly ferromagnetic metals. Guo \textit{et al}.
\cite{Guo14} predicted a significant magnetotransport effect in slightly
magnetized Pt. However, these calculations are for bulk materials and cannot
be used to model a monolayer-scale proximity effect on a magnetic insulator.
We are not aware of a realistic theoretical model for magnetotransport
dominated by the MPE.

Kikkawa \textit{et al}. \cite{Kikkawa13} addressed the possible contribution
of an MPE-induced anomalous Nernst effect (ANE) generated by an in-plane
temperature gradient over a Pt$|$YIG bilayer with an out-of-plane
magnetization. The observed Hall voltage was less than $5\%$ of that for an
in-plane magnetization and out-of-plane temperature gradient, i.e. the sum of
the ANE and the SSE, leading to the conclusion that an MPE-induced ANE may be disregarded.

The reported temperature dependence of the SMR is not unequivocal. Some groups
only find a decrease of the SMR amplitude at low temperatures (see
section~\ref{temperature}), others report an increasingly significant AMR
contribution in Pd$|$YIG \cite{Lin14}, Ta$|$YIG \cite{Yang14}, IrMn$|$YIG
\cite{XZhou15}, Pt$|$YIG \cite{XZhou15}, Pt$|$Co$_{2}$FeAl \cite{Zhang15}, and
Pt$|$LaCoO$_{3}$ \cite{Shang15} that might indicate a low temperature MPE.
Puzzling is the sign change of the AHE-like SMR at temperatures between a few
Kelvin and room temperature \cite{Qu14,Meyer15,XZhou15,Shang15}, suggesting
either a temperature-dependent $G_{i}$ or a new low-temperature transport
phenomenon, possibly caused by the MPE or interface disorder (discussed in
section~\ref{interface-roughness}). An AHE-like signal at low temperatures and
at magnetic fields higher than the saturation field of YIG \cite{Lin13,Miao14}
is interpreted as evidence for a hard magnetization texture of unknown
character at the interface. Further studies are required to shed light on the
physical mechanisms that may affect the MR at low temperatures. At room
temperature, the simple SMR model appears to be as yet unchallenged.

\subsection{Interface roughness and AHE-like SMR}

\label{interface-roughness}

Surface roughness scattering disregarded here affects many in-plane transport
properties of ultrathin metallic films. These effects can in principle be
modeled by semiclassical Boltzmann or diffusion theories
\cite{Fuchs38,Sondheimer52}. When the metallic films become thinner, their
resistivity increases by either reduced crystal quality or the increased
importance of surface roughness. Parameters such as the spin-flip diffusion
length and spin Hall angle may become thickness dependent as well. The AMR of
thin films of metallic ferromagnets is known to be affected by surface
roughness \cite{Rijks95,Rijks97}. The spin-dependent scattering at rough
interfaces is invoked to explain the current in-plane giant magnetoresistance
\cite{Camley89}. The SHE and the SMR in N$|$FI should also be affected by
roughness at the interface causing, for example, a spin-dependent mean-free
path in the normal metal even without an equilibrium MPE. Interface roughness
has minor effects on the spin-mixing conductance (when the SOC is small)
\cite{Brataas06}. Moreover, for out-of-plane magnetizations, spin-dependent
roughness can drive a transverse charge current by the ISHE, competing with
the contribution from the imaginary part of the spin-mixing conductance in (\ref{rho-2}).
While (\ref{rho-2}) is a second-order contribution of the SOC,
the contribution due to the spin-dependent roughness scales linearly to $\theta_{\mathrm{SH}}$ \cite{Chen14,SSZhang15-2}.
The two mechanisms can therefore be distinguished by measurements on materials with opposite signs of spin Hall angle such as Pt and Ta \cite{SSZhang15-2}.

Experimentally, the SMR in Pt$|$FM$|$MgO with (sub) monolayer ($0.3$ or
$0.6\,$nm) Co$_{2}$FeAl was explained in terms of an interference of
spin-dependent scattering at magnetic clusters and the SHE \cite{Zhang15}. A
quantitative analysis of the effects of roughness might be possible if induced
by controlled ion bombardment \cite{Miao13,Shan}.

\subsection{Interface spin-orbit couplings}

The SMR model attributes the spin polarization to the SHE, which requires the
SOC in the bulk layer, but disregards the SOC at the F$|$N interface. However,
interface SOC can modulate transport parameters that can be separated from
bulk effects experimentally only by tedious thickness-dependent studies. A
material-dependent theory of spin transport through F$|$N interfaces is not
available. However, an effective spin-mixing conductance can parameterize
computed results for the spin pumping contribution to the Gilbert damping and
interface spin flips \cite{Bass07} in N$|$Py$|$N sandwiches \cite{Liu14}. Such
an effective spin-mixing conductance appears to be not very sensitive to the
SOC for intermetallic interfaces such as Py$|$Pt \cite{Liu14}. 
A large spin-mixing conductance has been reported for Py$|$Cu$|$YIG \cite{Wang14}.
Experiments with non-local spin valves on a YIG substrated were interpreted in
terms of a finite Cu$|$YIG spin-mixing conductance; its relatively small value
might be caused by interface contamination \cite{Villamor15}. By a similar
technique, i.e. using a non-local Py$|$Al spin valves on a YIG substrate,
Dejene \textit{et al.} \cite{Dejene15} distill a large mixing conductance, but
the modulation as a function of YIG magnetization direction is smaller than
expected, which can be explained by thermal fluctuations
\cite{Cahaya14,Flipse14,Xiao15} (see section~\ref{temperature}).

The Edelstein effect refers to a current-induced spin accumulation with
polarization normal to the current direction in the plane of the
two-dimensional electron gas in which inversion symmetry is broken by a normal
electric field that can be described by the Rashba Hamiltonian
\cite{Bychkov84}. The Rashba Hamiltonian has been invoked also for thin
metallic films, in which non-equivalent interfaces are the source of the
inversion symmetry-breaking \cite{Manchon08,Manchon09}. An MR was predicted
for N$|$FI bilayers in the presence of a Rashba interaction at the interface
(and without SHE in the bulk) that depends on the magnetization direction just
like the SMR \cite{Grigoryan14}. Other theoretical studies of the MR in
ferromagnetic metallic thin films on a conventional insulating substrate with
Rashba SOC based on the Boltzmann equation \cite{Zhang14,SSZhang15} also
predict a contribution to the magnetoresistance which has the same
magnetization dependence as the SMR ($m_{y}^{2}$) but a different thickness
dependence \cite{SSZhang15}. Thus in principle, thickness-dependent studies
could help distinguish these different mechanisms.

Recent positron annihilation experiments observed current-induced spin
polarizations in normal metal films without proximity ferromagnets
\cite{Kawasuso13,HJZhang14}. The measured spin polarization is about one order
of magnitude larger than that estimated by the diffusive SHE model defined as
$\mu_{s}^{0}$ in section~\ref{theory-smr}. The authors attribute the
observations to the Edelstein effect \cite{Edelstein90}, i.e. a spin
accumulation with an in-plane polarization normal to the applied current
direction in a surface state, which may be interpreted as a two-dimensional
limit of the SHE. Jellium model calculations suggest a universal
current-induced spin polarization at metallic surfaces of $\left\vert \vec
{s}\right\vert =\Gamma j_{c}$ \cite{Tokatly15}, where $\Gamma$ is a material
dependent constant that appears to be large for Pt and Ta, but small for Au
and Cu \cite{HJZhang14}.

\section{Developments}

\label{developments}

We discuss in the following ongoing developments related to the SMR.

\subsection{Dynamics}

The SMR model took the magnetization in the FI to be static, which is not
generally the case since spin currents absorbed by the FI generate
spin-transfer torques which can amplify or attenuate \cite{Wang11} and even
generate spin waves \cite{Kajiwara10}. The threshold current for spin wave
excitations in N$|$FI is higher than found in the experiments even when
easy-axis magnetic surface anisotropy is taken into account
\cite{Xiao12,Xiao13,Zhou13}. Recent experiments show that current-induced
auto-oscillation in YIG can be achieved by current densities as predicteed,
but only when reducing the lateral dimensions of the FI \cite{Collet15}.

An ac SMR, or the spin Hall magnetoimpedance, was observed experimentally
\cite{Lotze14} to persists up to at least $4\,$GHz, allowing fast readouts of
the magnetization in N$|$FI bilayers. The SMR also allows observation of spin
torque induced FMR \cite{Chiba14}, i.e. the observation of a dc voltage by
down conversion of an applied ac current by the oscillating resistance under
resonance. Schreier \textit{et al}. \cite{Schreier14} and Sklenar \textit{et
al}. \cite{Sklenar15} separate the effects of the spin-transfer torque from
that of Oersted fields generated by the applied ac current and spin pumping
from the SMR rectification by comparing samples with different YIG and Pt
thickness \cite{Schreier14,Sklenar15}. This technique appears to be more
sensitive than applying microwaves via a coplanar wave guide \cite{Iguchi14}.

\subsection{Temperature dependence}

\label{temperature}

As a zero temperature theory, the SMR model described here assumes perfect
reflection of a spin current with spin polarization collinear to the
magnetization. At finite temperatures, the magnetization fluctuates and
thereby allows for a finite spin current even for collinear configurations
\cite{Cahaya14,Flipse14,Bender,Xiao15}. The SMR model thereby overestimates
the magnetoresistance. A larger than expected loss of spin accumulation in
collinear configurations has been inferred from experiments for YIG$|$Al
\cite{Dejene15}, which could imply that the spin-mixing conductances fitted to
room temperature SMR experiments are underestimated.

In bilayers of Pt, Ta, and Pd with YIG
\cite{Meyer14,Marmion14,Yang14,Lin13,Lin14}. SMR decrease with temperature
when going from $300\,$K to $10\,$K. This contradicts the (above) notion of a
spin current in the collinear configuration that is induced by thermal
fluctuations, which should lead to a larger SMR at lower temperatures. The
effect might be caused by a temperature dependence of the spin Hall angle
\cite{Meyer14}. A non-monotonic temperature dependence of the SMR in Pt$|$YIG
and Pd$|$YIG with a maximum at $100\,$K was explained in terms of a spin
diffusion length that is inversely proportional to the temperature, as
expected from the Elliot-Yafet impurity scattering model
\cite{Marmion14,Lin14}. In the regime from ambient to the Curie temperature of
YIG, the SMR decreases monotonically as expected \cite{Uchida15}. The
observation that the SMR decreases with temperature at a higher power than the
magnetization could be caused by a temperature-dependent spin-mixing
conductance \cite{Uchida15}.

Spin currents generated in a lateral F$|$N$_{1}|$F spin valve can be absorbed
by another contact (N$_{2}$) placed between the magnetic electrodes on the
path of the diffusive spin current in N$_{1}$. This technique gives access to
the spin diffusion length and spin Hall angle (via the ISHE) of N$_{2}$
\cite{Kimura07,Niimi12}, and was used to measure their temperature dependence
for Pt and Au \cite{Isasa15,Isasa15_err}. In Pt, both spin Hall angle and
electric resistivity are found to increase linearly with temp$\frac{{}}{{}}%
$erature, implying an intrinsic mechanism for the spin Hall conductivity
\cite{Sinova15}, while the SHE in Au appears to be dominated by extrinsic
effects \cite{Isasa15}. Such an analysis could be helpful to clarify the
microscopic mechanism responsible for the temperature dependence of the SMR.

The spin Seebeck effect in gadolinium (Gd) iron garnet (Gd$_{3}$Fe$_{5}%
$O$_{12}$) \cite{Gepraegs14} changes sign twice when the temperature is
lowered from $300\,$K to $35\,$K, and is explained in terms of different
contributions from the magnetic sublattices. Experimental results for the
temperature dependence of the SMR of rare earth iron garnets might help
establishing the importance of interface vs. bulk effects to explain this
intriguing phenomenon.

\subsection{Spin valves and superlattices}

The magnetizations in FI$|$N$|$FI spin valves can be coupled, in the ground
states by dipolar and non-local exchange interactions. A dynamic exchange
interaction is generated by spin current emission and absorption when
magnetizations are moving \cite{Heinrich11}. The magnetic state can be
detected by the SMR, which is enhanced when the magnetizations are collinear
and reduced when perpendicular to each other \cite{Chen13}. The SHE interferes
when the charge current through the spacer becomes appreciable. Applied
currents were found to modulate the damping of FI$|$N$|$FI spin valve and
$\left(  \text{FI$|$N}|\right)  _{n}$ superlattice dynamics. The critical
currents to excite magnetization dynamics (macrospin and spin waves) depend on
the relative orientations of magnetizations \cite{Skarsvag14}. The dynamic
coupling between spin waves due to spin pumping and spin-transfer torques is
in general mode-dependent \cite{Skarsvag14-2}.

\subsection{Bilayers of insulating and metallic magnets}

Recent studies address bilayers made of a metallic ferromagnet such as
permalloy and a ferromagnetic insulator such as YIG
\cite{Lu13-1,Miao13,Saitoh13,Tsukahara13,Wang14,Azevedo14,Hyde14}. The MR in
these systems \cite{Lu13-1,Miao13,Saitoh13} can be fully explained neither by
the conventional AMR nor by the SMR theory that only holds for paramagnetic
conductors. The MPE can induce a magnetic moment on a highly susceptible
paramagnet, but should have minor effects on a strong ferromagnet, although Lu
\textit{et al}. \cite{Lu13-1} argue otherwise. The Hall voltages induced by
thermal \cite{Miao13} and FMR \cite{Tsukahara13} spin pumping have been
interpreted in terms of the spin Hall angle of the ferromagnet, where
Tsukahara \textit{et al}. extracted a spin Hall angle of $\theta_{\mathrm{SH}%
}=0.5\%-1\%$ for Py. It was pointed out that dc Hall voltages can be generated
as well by spin rectification caused by the dynamics and the AMR in Py
\cite{Hyde14}. Spin pumping as detected by Hall voltages in W$|$Py$|$YIG
trilayers indicate opposite signs of the spin Hall angles in Py and W
\cite{Wang14}. However, the mechanism of spin pumping into a metallic
ferromagnet is not yet fully understood \cite{Azevedo14}, so estimates of the
spin Hall angle in magnetic metals should be taken with a grain of salt. By
decoupling Py and YIG with a copper spacer layer \cite{Wang14}, the problem
becomes amenable to the conventional theory, however, a $\theta_{\mathrm{SH}%
}=2\%$ for Py was deduced. First principles calculations predict extrinsic
spin Hall angles in diluted ferromagnetic alloys of $\theta_{\mathrm{SH}}%
\geq1\%$ \cite{Zimmermann14}.

\section{Conclusions/outlook}

\label{conclusions}

The discovery of a magnetoresistance in nominally non-magnetic metals next to
magnetic insulators has stimulated many subsequent studies. Experiments from
different groups and sample growth techniques yield consistent results that
are well reproduced by a few-parameter, simple and intuitive model called
\textquotedblleft spin Hall magnetoresistance (SMR)\textquotedblright\ that
explains the observations by the concerted action of the direct and inverse
spin Hall effects. While providing a convenient parametrization scheme, the
detailed physics and material dependence is still a matter of discussion. By
using an exchange-only theory of spin-transfer torque, the effect of
interfacial spin-orbit coupling (SOC) is disregarded. This does not
necessarily reduce the quality of the fits, but the parameters must then be
considered to represent an effective average of interface and bulk
contributions with the same symmetry. Fitting parameters that depend strongly
on the thickness of the normal metal film have not been reported to date.
First-principles calculations including the SOC and disorder such as those by
Liu \textit{et al}. \cite{Liu14} should be able to shed more light on this
issue. Measurements at low temperatures and high magnetic fields have been
interpreted in terms of a magnetic proximity effect
\cite{Lin13,Lin14,Yang14,Qu14,XZhou15,Zhang15,Shang15,Meyer15,Miao14},
although even a qualitative theory in support of this conjecture is lacking.
XMCD experiments that in principle provide a definite answer lead to
contradicting results \cite{Geprags12,Lu13}. Also here first-principles
calculation may importantly help to not only estimate the magnitude and
penetration of the induced moments, but also its effects on transport and
taking into account the magnetic fluctuations induced by finite temperatures
\cite{Liu15}.

The SMR has been employed to study spin-correlations in antiferromagnets
\cite{Han14} and multiferroic materials \cite{Aisha}. In these studies a
magnetic field has to be applied in order to observe a finite SMR, which
implies that an interface magnetic susceptibility is sampled, rather than a
ground state spin texture. While the SMR is a sensitive detector for subtle
spin correlations \cite{Aisha}, more work is needed to better understand the
implications. The spin Seebeck effect in paramagnets under strong magnetic
fields \cite{Wu15} should be accompanied by an SMR. The SMR has been observed
in the current-direction dependence of the Johnson-Nyquist noise of YIG$\vert$Pt bilayers, 
which can be understood in terms of the fluctuation dissipation
theorem \cite{Kamra14}. Another direction worth exploring are thermal
analogies of the SMR \cite{Bauer12}: we may, e.g., expect a dependence of the
in-plane thermopower on the magnetization direction in N$|$FI bilayers as a
fingerprint of the spin Nernst effect. Very recently, a non-local SMR (or
magnon-drag effect \cite{Zhang12}) was observed in Pt$\vert$YIG$\vert$
Pt lateral heterostructures \cite{Cornelissen,Goennenwein}, in which a current
in one contact generates a voltage in the other contact that is maximal when
both are normal to the magnetization direction. 


\ack This work was supported by a Grant-in-Aid for Scientific Research of the
Japan Society for the Promotion of Science, Grant Nos.~26103006, 22540346,
25247056, 25220910, 268063, 26790037, the FOM (Stichting voor Fundamenteel
Onderzoek der Materie), EU-ICT-7 {InSpin,}\ the ICC-IMR and E-IMR,
the Casio Science Promotion Foundation,
and DFG Priority Programme 1538 \textquotedblleft{Spin-Caloric Transport}%
\textquotedblright\ (GO 944/4, BA 2954/2).





\end{document}